\documentstyle[psfig,conf_iap,]{article}
\begin{document}
\heading{%
%
The Hot Phase in the Galactic Disk
%
} 
\par\medskip\noindent
\author{%
Edward B. Jenkins$^{1}$, David V. Bowen$^{1}$, Kenneth R.
Sembach$^{2}$,\\
FUSE Science Team
}
\address{%
Princeton University Observatory, Princeton, NJ 08544-1001, USA
}
\address{Dept. of Physics \& Astronomy, Johns Hopkins University,
Baltimore, MD 21218-2686, USA}

\begin{abstract}
Spectra of $\sim 150$ early-type stars in the disk of the Milky Way
recorded by FUSE reveal absorption features from interstellar O~VI, a
tracer of collisionally ionized gases at temperatures near 300,000 K. 
(Hotter material is better sensed by its diffuse x-ray emission.)  The
FUSE survey indicates that hot gas in the plane of the Galaxy yields an
average density $n({\rm O~VI}) = 1.7\times 10^{-8}\,{\rm cm}^{-3}$, a
value consistent with earlier results from the {\it Copernicus\/}
satellite, but at great distances there is a large dispersion in the
measurements from one line of sight to another.  This dispersion
indicates that O~VI absorbing regions probably have a broad distribution
of sizes, perhaps giving a power-law behavior for column densities. 
Comparisons of O~VI with other Li-like ions should give some guidance on
the nature and origin of the O~VI-absorbing regions, but poor
correspondences in the velocity profile shapes will complicate the
interpretations.

\end{abstract}
\section{Principal Ways to Detect Hot Gas}
An important achievement of space astronomy has been the discovery of a
hot phase of the interstellar medium, confirming an early prediction by
Spitzer  [1] that this phase may be pervasive -- a conclusion based on
the need to explain the confinement of cool clouds in the Galactic halo. 
We now know that highly ionized gases spanning the temperature range
from $10^5$ to $10^7\,$K are indeed an important constituent of the ISM,
in both the plane and halo of the Galaxy.

There are three fundamental ways to detect this hot gas:
\begin{enumerate}
\item Absorption by highly ionized atoms in the UV spectra of background
sources (stars for Galactic disk gas, quasars or AGN for the Galactic
halo or IGM)
\item Emision of soft x-rays over broad energy bands, with the radiation
coming from collisionally excited levels, free-free emission, or
recombination
\item Emission of discrete UV lines from collisionally excited levels
\end{enumerate}
Most of the findings on the hot phase have come from the phenomena 1 and
2 above; item 3 in this list is technically difficult to observe, and so
far the results have been rather primitive and limited in scope.

\begin{figure}
\centerline{\vbox{
\psfig{figure=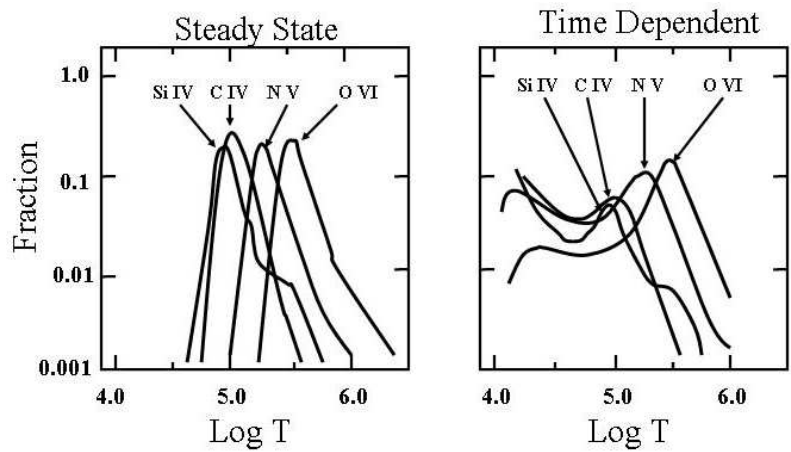,height=6.cm}
}}
\caption[]{Logarithmic plots of computed fractional abundances  [2] as a
function of plasma temperature for the principal ions that can be seen
in absorption, assuming the gas is in thermal equilibrium (left-hand
panel) and that the gas is cooling radiatively at constant volume
(right-hand panel).\label{ioniz_frac}}
\end{figure}

The ions Si~IV, C~IV, N~V and O~VI have strong transitions from their
ground states in the ultraviolet, and they are the most suitable species
for revealing the presence of hot gas through their absorption features. 
Figure~\ref{ioniz_frac} shows the abundances of these ions relative to
the total abundances of the respective elements for different plasma
temperatures  [2].  In collisional ionization equilibrium, the ion
fraction peaks are well confined to narrow temperature ranges (left-hand
panel of the figure).  For gas that is cooling radiatively, the cooling
time is shorter than the recombination times and thus the ion fraction
peaks are strung out toward lower temperatures (right-hand panel).  This
property, plus the chance that photoionizing fluxes near hot stars could
produce measurable amounts of Si~IV and C~IV  [3], indicates that the
presence of these species does not unequivocally indicate the existence
of gas with $T\geq 10^5\,$K.

The most useful ion to observe is O~VI, which has a strong resonance
doublet at $\lambda\lambda$ 1031.9, 1037.6.  The ionization potential of
${\rm O}^{+4}$ is 114~eV, which is more energetic than most of the
photons emitted by the hottest stars, except for hydrogen-rich white
dwarfs  [4].  Thus, under most circumstances O~VI is unlikely to arise
from photoionization.

Observations of O~VI and the soft x-ray background have provided our
most detailed look at the character of the hot gas and its distribution
in space.  By chance, these two research industries started
simultaneously.  As reported in a single issue of ApJ Letters in 1974,
investigators of the soft x-ray background  [5] and UV astronomers 
[6,7] independently arrived at the conclusion that their respective
results supported the existence of a significant hot phase for the ISM
in the Galactic disk.  Several years later Jenkins  [8] published his
survey of O~VI absorption toward 72 stars at distances up to about 2 kpc
from the Sun.   A detailed map of the x-ray sky at several energies has
now emerged from an extensive survey of diffuse emission using Rosat 
[9].  The O~VI and x-ray samples sense the hot gas in very different
ways (as outlined in Table~1 below), so the two efforts complement each
other.  A sensitive spectrograph on the ORFEUS mission extended our
reach for detecting O~VI absorption to stars in the Galactic halo 
[10,11], and this research was followed up by a more extensive survey of
the halo using FUSE   [12] (see also an article by Savage et al. in this
volume).  In the pre-FUSE era, the only report on UV line emission from
hot gas (outside of supernova remnants) came from Martin \& Bowyer  [13]
(for recent FUSE results, see the article by Shelton in this volume).

\begin{center}
\begin{small}
\begin{tabular}{l l l}
\multicolumn{3}{l}{{\bf Table~1.} General Properties,
Strengths/Weaknesses of Different Observations } \\
\hline
\multicolumn{1}{l}{}&\multicolumn{1}{l}{Soft x-ray emission}
&\multicolumn{1}{l}{O~VI absorption}\\
\hline
\\
\multicolumn{1}{l}{Quantity measured}
&\multicolumn{1}{l}{$\int n(e)^2dl$}
&\multicolumn{1}{l}{$\int n(e)dl$}\\
\\
\multicolumn{1}{l}{Characteristic}
&\multicolumn{1}{l}{$\log T \geq 5.8$}
&\multicolumn{1}{l}{$5.3 < \log T < 5.7$}\\
\multicolumn{1}{l}{temperatures}\\
\\
\multicolumn{1}{l}{Effect of (low}
&\multicolumn{1}{l}{Energy-dependent}
&\multicolumn{1}{l}{Target's overall UV}\\
\multicolumn{1}{l}{temperature) ISM}
&\multicolumn{1}{l}{attenuation by}
&\multicolumn{1}{l}{brightness attenuated by}\\
\multicolumn{1}{l}{}
&\multicolumn{1}{l}{foreground or}
&\multicolumn{1}{l}{dust, otherwise no effect}\\
\multicolumn{1}{l}{}&\multicolumn{1}{l}{interspersed gas}\\
\\
\multicolumn{1}{l}{Available sampling}
&\multicolumn{1}{l}{Map whole sky, path}
&\multicolumn{1}{l}{Only sample toward stars,}\\
\multicolumn{1}{l}{}
&\multicolumn{1}{l}{length indefinite}
&\multicolumn{1}{l}{path length definite (except}\\
\multicolumn{1}{l}{}
&\multicolumn{1}{l}{}
&\multicolumn{1}{l}{for extragalactic targets)}\\
\\
\multicolumn{1}{l}{Radial velocities}
&\multicolumn{1}{l}{Not measurable}
&\multicolumn{1}{l}{Can measure}\\
\multicolumn{1}{l}{and their dispersion}
&\multicolumn{1}{l}{}
&\multicolumn{1}{l}{centroids and widths}\\
\\
\hline
\end{tabular}
\end{small}
\end{center}

\section{A New Survey of O~VI in the Galactic Disk}

Even though the results summarized above gave important insights on the
hot gas, nagging problems remained.  The x-ray observers arrived at the
conclusion that the Sun is immersed in a large volume, extending out to
about 100~pc, containing mostly hot gas.  How much are the conclusions
about the general nature of the gas from the early O~VI survey skewed by
this ever-present contribution  [14]?  Does the average density of O~VI
change if we look to greater distances?  To answer these (and other)
questions, the FUSE team undertook a survey of about 150 stars in the
Galactic plane.  Preliminary conclusions from this new survey will
follow.

The interstellar medium is heated by interstellar shocks.  In a region
behind a shock front, before the gas has had a chance to cool, the
temperature $T$ is given by the formula
\begin{equation}\label{postshock_T}
T=3\mu v^2/16k=1.4\times 10^5v_{100}^2\,{\rm K}
\end{equation}
where $v_{100}$ is the shock velocity in units of $100\,{\rm km~s}^{-1}$
($\mu$ is the mean particle mass and $k$ is the Boltzmann constant).  We
recall from Fig.~\ref{ioniz_frac} that O~VI reaches its maximum
concentration at $\log T=5.5$, hence a shock having a velocity of order
$150\,{\rm km~s}^{-1}$ is needed to heat the gas to a sufficient
temperature to produce O~VI.  The FUSE observation of HD~75309 behind
the Vela supernova remnant shows a very strong O~VI feature at about
$-100\,{\rm km~s}^{-1}$ (with a plausible projection factor, this could
well represent a shock with $v_{100}=1.5$), and evidence for high
velocity O~VI has been seen toward Vela in an earlier study using {\it
Copernicus\/}  [15].  However, for other directions in the sky, the O~VI
features are within a few tens of ${\rm km~s}^{-1}$ of either the local
standard of rest or the expected velocity of the medium near the target
star.  A typical width of such a feature is about $42\,{\rm km~s}^{-1}$
(FWHM, after correcting for instrumental smearing), which is not much
larger than the expected thermal Doppler broadening width (FWHM) of
$30\,{\rm km~s}^{-1}$ at $T=10^{5.5}\,$K.  Thus, any explanation for the
origin of the O~VI material must acknowledge the fact that the gas has
somehow been stabilized in some fashion, or that what we see is a
secondary phenomenon that is kinematically coupled to the
low-temperature ISM.

\begin{figure}
\centerline{\vbox{
\psfig{figure=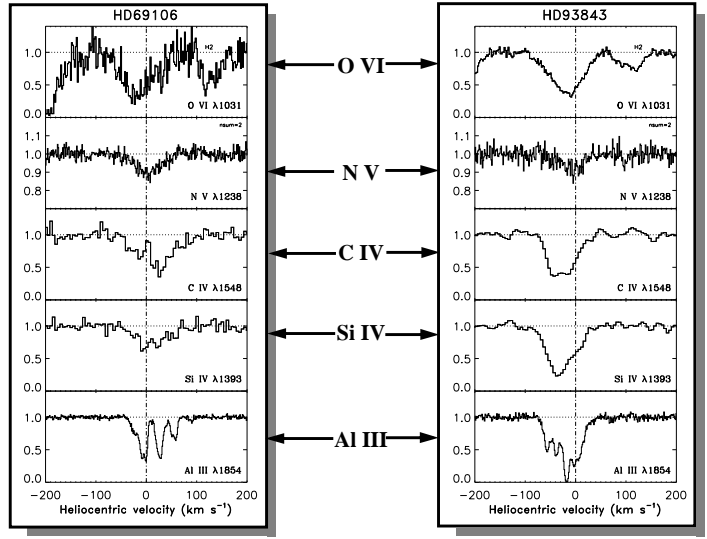,height=7.5cm,angle=270}
}}
\caption[]{Absorption profiles of highly ionized Li-like species toward
HD~69106 and HD~93843.  The profile of Al~III has been added to indicate
the probable behavior of photoioinized gas.\label{different_ions}}
\end{figure}

It is possible that most of the hot gas in the disk of the Galaxy is not
visible in O~VI, either because it is too hot (i.e., $T>10^6\,$K) or has
such a large velocity dispersion that the O~VI absorption is not
recognizable because the features are lost in the undulating continuum
of the target star.  This may explain why there seems to be not enough
O~VI compared to the soft x-ray emission, unless the filling factor of
the hot gas is extraordinarily small and the thermal pressures large 
[16] (recall from Table~1 that ${\rm O~VI}\Rightarrow n(e)$, x-rays
$\Rightarrow n(e)^2$).  Perhaps the O~VI profiles arise from conductive
interfaces between cool clouds and a much hotter medium  [17] or in
turbulent mixing layers on the surfaces of clouds that are moving with
respect to their surroundings  [18].  One way to resolve the question is
to compare O~VI to the other highly ionized species (see
Fig.~\ref{ioniz_frac}) and then determine which of the theoretical
models show the best consistency with the observations  [19]. 
Unfortunately, this might not be such an easy task because the velocity
profiles of the different species do not match each other very
well, as is demonstrated for two cases in Fig.~\ref{different_ions}. 
This nonconformity indicates that simple comparisons of total column
densities for different ions will not be appropriate.

\begin{figure}
\centerline{\vbox{
\psfig{figure=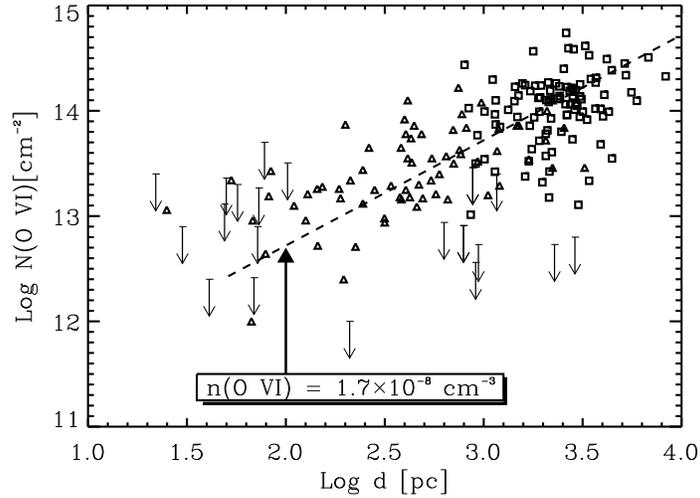,height=7.cm}
}}
\caption[]{The relationship between $N({\rm O~VI})$ and distance.  Boxes
represent data obtained from the FUSE disk survey, while triangles
indicate measurements from the {\it Copernicus\/} survey  [8] and a
separate study of the Local Bubble using FUSE.\label{N_vs_d}}
\end{figure}

We turn next to the information conveyed by the column densities of O~VI
and the respective distances to the targets.  An important theme in the
interpretation of the {\it Copernicus\/} survey by Jenkins  [20] was
that if O~VI clouds were of uniform size and distributed at random,
their number per unit volume could be estimated by the magnitudes of
fluctuations in the O~VI column per unit distance caused by chance encounters. 
An important test of that picture is to see if the relative magnitudes
of these fluctuations decrease at greater distances.  The points in
Fig.~\ref{N_vs_d} indicate that this is not happening.  We are
forced to discredit the idea that the O~VI-bearing regions consist of
uniform packets -- instead there is probably a broad power-law type
distribution in sizes.  As we go to greater distances, very large O~VI
regions begin to enter the picture and cause fluctuations that start to
overwhelm the effects of smaller regions seen over short distances.

Finally, we address the question about the overall filling factor $f$ of
the O~VI regions.  Over a particular sight-line distance $d$, random
interceptions of O~VI regions occupying a relative volume $f$ is
equivalent to viewing a single region of thickness $fd$.  The thermal
pressure of that region (divided by the Boltzmann constant $k$) is given
by
\begin{equation}\label{filling_factor}
{p\over k} = n_{\rm total}T = {N({\rm O~VI})\over fd}~\left({n_{\rm
O}\over n_{\rm O~VI}}\right)_T~{n_{\rm H}\over n_{\rm O}} ~ {n_{\rm
total}\over n_{\rm H}} ~ T
\end{equation}
where we adopt reasonable numbers for the last four terms: the ion
fraction $(n_{\rm O}/n_{\rm O~VI})_T\simeq 10$ (but it could be larger
if $T$ is very different from $10^{5.5}\,$K), the ratio of hydrogen to
oxygen $n_{\rm H}/n_{\rm O}$ equals the cosmic abundance ratio 3100, and
the ratio of all particles to protons $n_{\rm total}/n_{\rm H}=2.3$.  If
we adopt the median value $N({\rm O~VI})/fd=1.7\times 10^{-8}\,{\rm
cm}^{-3}$ (see Fig.~\ref{N_vs_d}) and solve for $f$ in
Eq.~\ref{filling_factor}, we obtain $f\geq 0.013$ if $p/k\leq 3\times
10^4\,{\rm cm}^{-3},\,$K and $T=10^{5.5}\,$K.

\acknowledgements{The new O~VI results were obtained for the Guaranteed
Time Team by the NASA-CNES FUSE mission operated by Johns Hopkins
University.  Financial support to US participants has been provided by
NASA contract NAS5-32985 and support for writing this paper came from
Contract 2440-60014 to Princeton University.}

\vfill

\begin{iapbib}{99}{
\bibitem{}Spitzer, L. 1956, \apj 124, 20

\bibitem{}Shapiro, P. R., \& Moore, R. T. 1976, \apj 207, 460

\bibitem{}Cowie, L. L., Taylor, W., \& York, D. G. 1981, \apj 248, 528

\bibitem{}Dupree, A. K., \& Raymond, J. C. 1983, \apj 275, L71

\bibitem{}Williamson, F. O., Sanders, W. T., Kraushaar, W. L., 
McCammon, D., Borken, R., \& Bunner, A. N. 1974, \apj 193, L133

\bibitem{}Jenkins, E. B., \& Meloy, D. A. 1974, \apj 193, L121

\bibitem{}York, D. G. 1974, \apj 193, L127

\bibitem{}Jenkins, E. B. 1978, \apj 219, 845

\bibitem{}Snowden, S. L., Egger, R., Freyberg, M. J., McCammon, D., 
Plucinsky, P. P., Sanders, W. T., Schmitt, J. H. M. M., Tr\"umper, J.,
\& Voges, W. 1997, \apj 485, 125

\bibitem{} Hurwitz, M., Bowyer, S., Kudritzki, R. P., \& Lennon, D. J. 
1995, \apj 450, 149

\bibitem{} Hurwitz, M., \& Bowyer, S. 1996, \apj 465, 296

\bibitem{} Savage, B. D., Sembach, K. R., Jenkins, E. B., Shull, J. M., 
York, D. G., Sonneborn, G., Moos, H. W., Friedman, S. D., Green, J. C., 
Oegerle, W. R., Blair, W. P., Kruk, J. W., \& Murphy, E. M. 2000, \apj
538, L27

\bibitem{} Martin, C., \& Bowyer, S. 1990, \apj 350, 242

\bibitem{} Shelton, R. L., \& Cox, D. P. 1994, \apj 434, 599

\bibitem{} Jenkins, E. B., Silk, J., \& Wallerstein, G. 1976, \apj
Suppl. 32, 681

\bibitem{} Shapiro, P. R., \& Field, G. B. 1976, \apj 205, 762

\bibitem{} Cowie, L. L., Jenkins, E. B., Songaila, A., \& York, D. G. 
1979, \apj 232, 467

\bibitem{} Slavin, J. D., Shull, J. M., \& Begelman, M. C. 1993, \apj 
407, 83

\bibitem{} Spitzer, L. 1996, \apj 458, L29

\bibitem{} Jenkins, E. B. 1978, \apj 220, 107




}
\end{iapbib}
\end{document}